\documentclass[conference]{IEEEtran}
\IEEEoverridecommandlockouts
\usepackage{cite}
\usepackage{amsmath,amssymb,amsfonts}
\usepackage{algorithmic}
\usepackage[table,xcdraw]{xcolor}
\usepackage{graphicx}
\usepackage{lipsum}
\usepackage{hyperref}
\usepackage{textcomp}
\ifCLASSOPTIONcompsoc
    \usepackage[caption=false, font=normalsize, labelfont=sf, textfont=sf]{subfig}
\else
\usepackage[caption=false, font=footnotesize]{subfig}
\fi
\usepackage{xcolor}
\def\BibTeX{{\rm B\kern-.05em{\sc i\kern-.025em b}\kern-.08em
    T\kern-.1667em\lower.7ex\hbox{E}\kern-.125emX}}
\begin{document}

\title{A Real-Time Retinomorphic Simulator Using a Conductance-Based Discrete Neuronal Network\\
\thanks{J. K. Eshraghian's contribution was supported by the Endeavour Research Leadership Award from the Australian Government, and iDataMap Corporation Pty. Ltd.}
}




\author{Seungbum Baek$^{1,5}$, Jason~K.~Eshraghian$^{2,5}$, Wesley Thio$^2$, Yulia Sandamirskaya$^3$, Herbert H.C. Iu $^4$ and Wei D. Lu$^2$\\
\IEEEauthorblockA{$^1$\textit{College of Electrical and Computer Engineering, Chungbuk National University, Cheongju 362763, South Korea}}
\IEEEauthorblockA{$^2$\textit{School of Electrical, Electronic and Computer Engineering, University of Michigan, Ann Arbor, MI 48109 USA}}
\IEEEauthorblockA{$^3$\textit{Institute of Neuroinformatics Neuroscience Center Zurich University and ETH Zurich, Switzerland}}
\IEEEauthorblockA{$^4$\textit{School of Electrical, Electronic and Computer Engineering, University of Western Australia, Crawley, WA 6009, Australia}}
\IEEEauthorblockA{$^5$\textit{These authors contributed equally to this manuscript.}}}

\maketitle
\begin{abstract}
We present an optimized conductance-based retina microcircuit simulator which transforms light stimuli into a series of graded and spiking action potentials through photo transduction. We use discrete retinal neuron blocks based on a collation of single-compartment models and morphologically realistic formulations, and successfully achieve a biologically real-time simulator. This is done by optimizing the numerical methods employed to solve the system of over 270 nonlinear ordinary differential equations and parameters. Our simulator includes some of the most recent advances in compartmental modeling to include five intrinsic ion currents of each cell whilst ensuring real-time performance, in attaining the ion-current and membrane responses of the photoreceptor rod and cone cells, the bipolar and amacrine cells, their laterally connected electrical and chemical synapses, and the output ganglion cell. It exhibits dynamical retinal behavior such as spike-frequency adaptation, rebound activation, fast-spiking, and subthreshold responsivity. Light stimuli incident at the photoreceptor rod and cone cells is modulated through the system of differential equations, enabling the user to probe the neuronal response at any point in the network. This is in contrast to many other retina encoding schemes which prefer to `black-box' the preceding stages to the spike train output. Our simulator is made available open source, with the hope that it will benefit neuroscientists and machine learning practitioners in better understanding the retina sub-circuitries, how retina cells optimize the representation of visual information, and in generating large datasets of biologically accurate graded and spiking responses. 

\end{abstract}

\section{Introduction}
This paper presents a retinal simulation platform that integrates the image processing that takes place within the vertebrate vision system, transforming incoming light into a spike train sent to the brain for interpretation. Retinal architecture is largely well understood, from the constituent cell types to their connectivity, but the manner in which retina sub-circuits perform computation has not yet been fully discerned \cite{Understanding2014}. Therefore, bridging the gap between biological plausibility and functional models are far from maturation. This is a crucial step to the realization of effective retinal prosthesis \cite{Optimization2019}, and fabricating high-performance image sensors that are on par with the specifications of the retina in terms of power dissipation, dynamic range, and resolution \cite{Lichsteiner2008, Eshraghian2020, Barranco2019, Moini1997, Eshraghian2018}. At present, most bio-inspired image processors trade-off the ability to pass low-frequency content and low spatial resolution, in favor of practicality.

Beyond hardware, a similar distinction exists between biological plausibility and functionality in retina modeling. At the microscale, a single neuron can be modeled by integrating electrophysiological current and voltage-clamp recordings into a mechanistic understanding of the neuron properties. This process can be represented as a single-compartmental model \cite{Hodgkin1952, Fohlmeister1990, Fohlmeister1997, Kameneva2011, Boinagrov2012}, a morphologically realistic approach \cite{Fohlmeister2010, Publio2012}, or using a series of block-compartments \cite{Schiefer2006, Werginz2014}. Single-compartment models use capacitance to mimic the bilipid membrane, in parallel with nonlinear conductances which act as the various transmembrane ion channels. Morphologically realistic models hone in on the physical features of a biological neuron, such as the soma, axon hillock, axon initial segment, and dendrites, which provide an improved, but structurally complex, approximation over single-compartments. Block-compartment models sacrifice biological plausibility for computational efficiency through pruning, and retain only the most necessary anatomical information from morphological models. These are the most common methods to represent single neurons.

\begin{figure}[t]
\centerline{\includegraphics[scale=0.6]{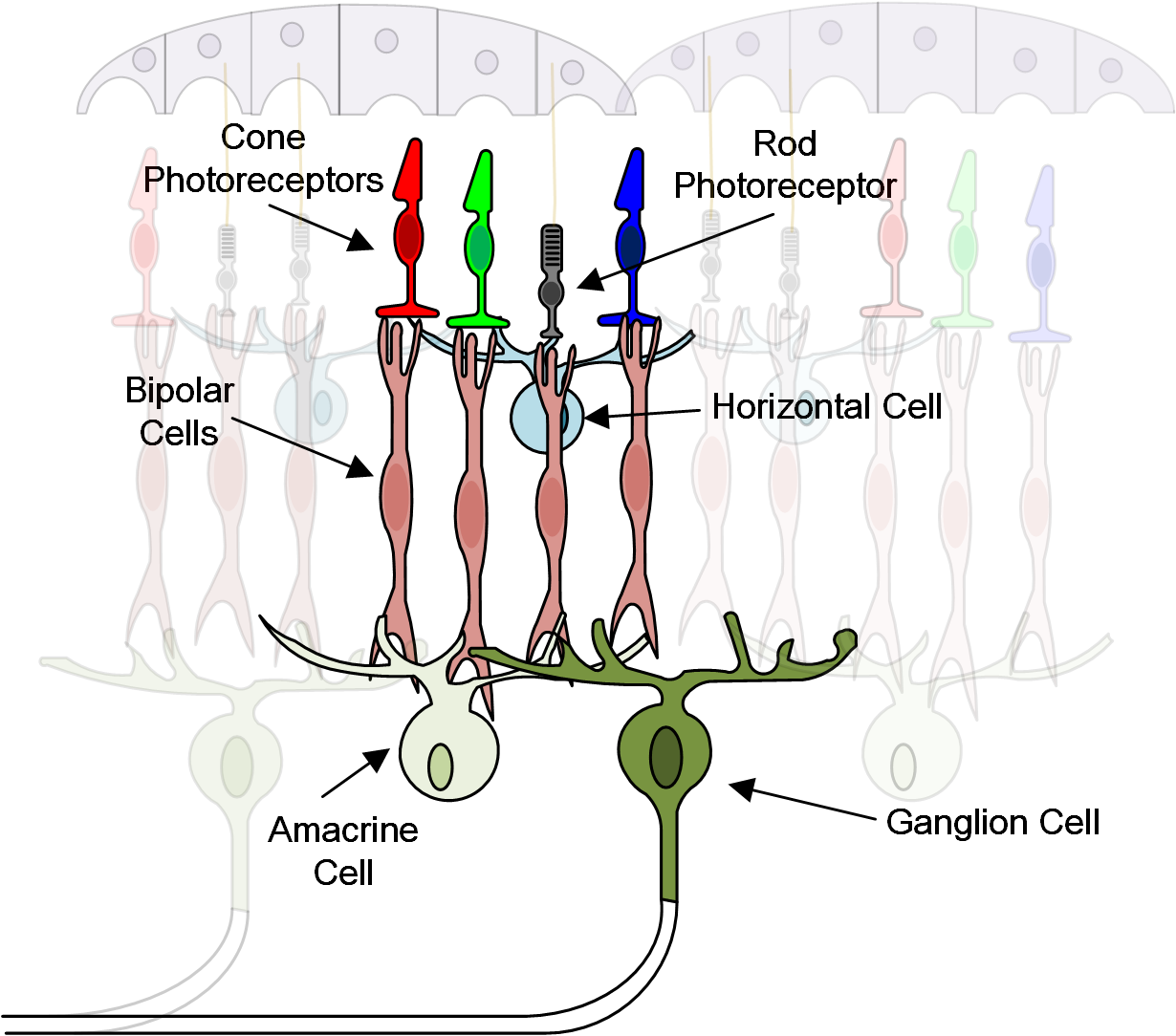}}
\caption{Dual rod and cone pathway. Light signals are transformed into electrical signals that flow from rod and cone cells through the individual pathways, and are integrated at the ganglion cell resulting in a spike train.}
\label{fig1}
\end{figure}

On a larger scale, retinal neurons will encode visual information as either a graded or spiking potential, and transfer the results to downstream neurons (in addition to lateral and feedback connections), as depicted in Fig.~\ref{fig1}. One approach to representing the neuronal network within the retina is to cascade individual neurons, each of which are capable of independent execution, and tie them together with computational models of synaptic connections. These types of cascaded models are mechanistically detailed, but each neuron adds an additional layer of computational cost. To achieve functional and real-time retina models, many morphological details of the neurons are omitted \cite{Eshraghian2016}. Interestingly, large-scale retina models that integrate morphologically accurate neurons are under-represented in the literature. This arises for a variety of reasons: electrophysiologists that conduct experimental research on the retina are not typically interested in large-scale models that unify well-understood results. Computational and cognitive neuroscientists who require datasets of action potentials often make severe simplifications for ease of processing in higher cortical areas.

The primary goal with this retina simulator is to provide a simple to use and intuitive discrete neuronal network model, that uses single-compartment and morphologically plausible neurons, coupled with their associated synapses to ensure biophysical accuracy with real-time processing. We achieve this by optimizing the numerical methods used to solve the large-scale system of nonlinear ordinary differential equations that represent the retina. By opening up access to this simulator, we simplify the process for interdisciplinary researchers, in particular computational neuroscientists and machine learning practitioners, to gain better insight to the retina's representation of visual information, to be able to generate large datasets of neuronal data that is otherwise difficult to procure, and to foster online bio-hybrid experimentation.




\section{Retina Model}
\begin{figure}[t]
\centerline{\includegraphics[scale=0.6]{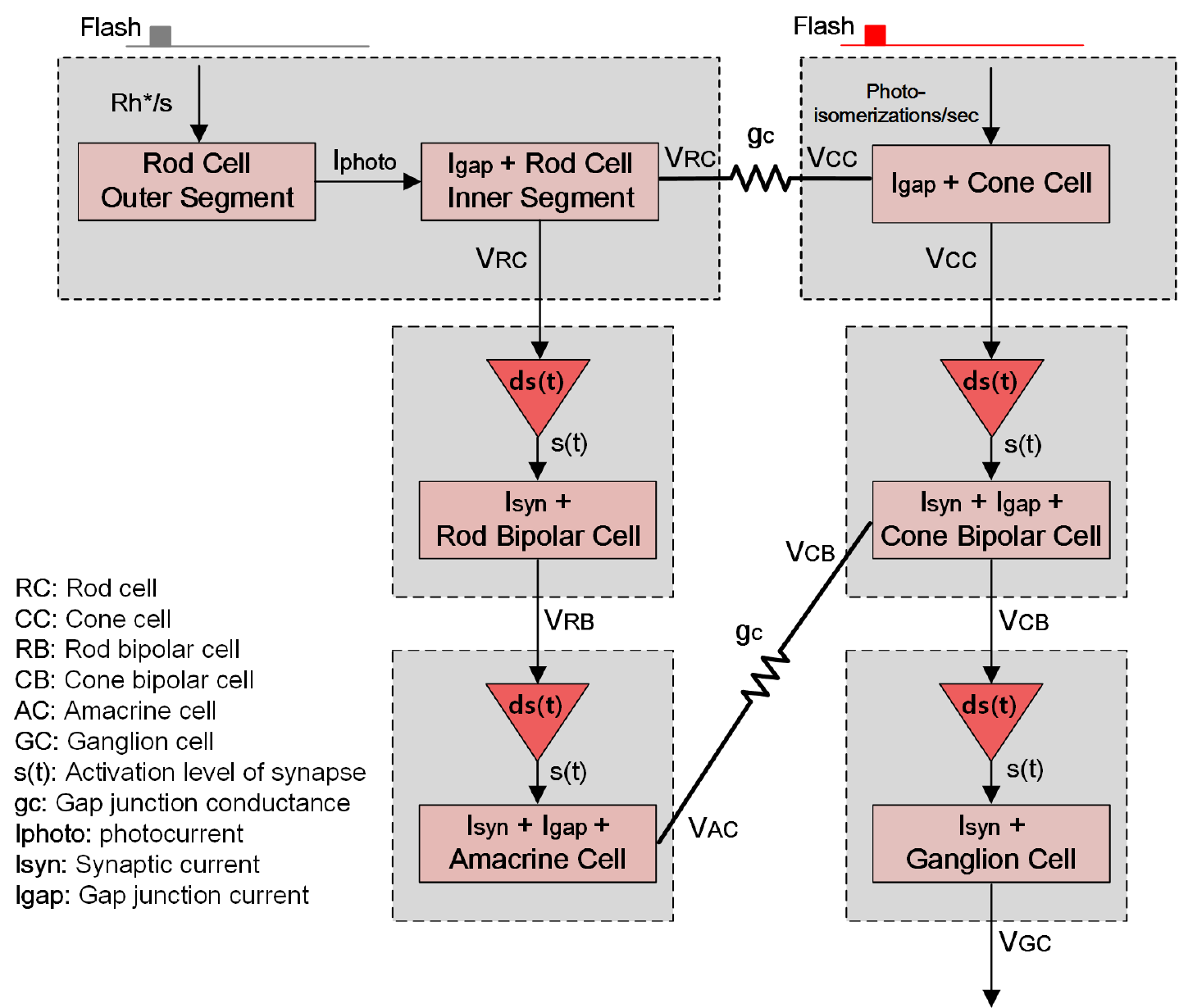}}
\caption{Schematic of the parallel signal flow pathway.}
\label{fig2}
\end{figure}

A high-level schematic of the retinal signal flow pathway is shown in Fig.~\ref{fig2}. The nonlinear system of ordinary differential equations we use in our model can be generalized by:

\begin{equation} \label{eq1}
C_m \frac{dV}{dt} = \sum I_{\rm ion}(V, \chi ; [\rm Ca^{2+}])+I_{syn}(V),
\end{equation}

\begin{equation} \label{eq2}
\frac{d[Ca^{2+}]}{dt} = I_{\rm Ca}(V)+G(V;[\rm Ca^{2+}]),
\end{equation}

\begin{equation} \label{eq3}
\frac{d\chi}{dt} = (1-\chi)\alpha_{\chi}(V)-\beta_{\chi}(V)\chi.
\end{equation}

\noindent $C_m$ and $V$ are membrane capacitance and potential, respectively; $I_{syn}$ is input synaptic current; $I_{\rm ion}$ is the sum of all ion channel currents; [Ca$^{2+}$] is the intracellular concentration of free calcium; $I_{\rm Ca}(V)$ represents all voltage-activated calcium currents; $G$ covers all other voltage and calcium-dependent processes; $\chi$ is the gating variable, and the forward-backwards reaction rates are $\alpha_{\chi}$ and $\beta_{\chi}$, respectively. The complete system of equations and initial conditions are subsumed within (\ref{eq1})--(\ref{eq3}), and the list can be found in the online appendix in \cite{appendix}. There are a total of 272 nonlinear differential equations and parameters, with a model collation summary provided in Table~I.

\begin{table}[]\scriptsize
\centering\caption{Discrete neuronal model blocks}
\begin{tabular}{
>{\columncolor[HTML]{FFFFFF}}l 
>{\columncolor[HTML]{FFFFFF}}l}
\hline\vspace{-6pt}
 &   \\ \hline
\multicolumn{1}{|c|}{\cellcolor[HTML]{FFFFFF}\textbf{Function}} & \multicolumn{1}{c|}{\cellcolor[HTML]{FFFFFF}\textbf{Literature}}\\ \hline
\multicolumn{1}{|c|}{\cellcolor[HTML]{FFFFFF}Phototransduction to spike train conversion} & \multicolumn{1}{c|}{\cellcolor[HTML]{FFFFFF}Eshraghian \textit{et al.} (2018)} \\
 \multicolumn{1}{|c|}{\cellcolor[HTML]{FFFFFF}Bandpass filtering of rod photoreceptor network} & \multicolumn{1}{c|}{\cellcolor[HTML]{FFFFFF}Kamiyama \textit{et al.} (2009)} \\
\multicolumn{1}{|c|}{\cellcolor[HTML]{FFFFFF}Ionic current in rod photoreceptor network} & \multicolumn{1}{c|}{\cellcolor[HTML]{FFFFFF}Kamiyama \textit{et al.}(1996)} \\
\multicolumn{1}{|c|}{\cellcolor[HTML]{FFFFFF}Voltage- and calcium-activated current of rod cell} & \multicolumn{1}{c|}{\cellcolor[HTML]{FFFFFF}Bader \textit{et al.}(1982)} \\
 \multicolumn{1}{|c|}{\cellcolor[HTML]{FFFFFF}Calcium and Chloride currents in cone cells} & \multicolumn{1}{c|}{\cellcolor[HTML]{FFFFFF}Maricq \textit{et al.} (1988)} \\ 
\multicolumn{1}{|c|}{\cellcolor[HTML]{FFFFFF}Ion channels of cone cells} & \multicolumn{1}{c|}{\cellcolor[HTML]{FFFFFF}Barnes \textit{et al.} (1989)} \\
\multicolumn{1}{|c|}{\cellcolor[HTML]{FFFFFF}Phototransduction in rod cells} & \multicolumn{1}{c|}{\cellcolor[HTML]{FFFFFF}Torre \textit{et al.} (1990)} \\ 
\multicolumn{1}{|c|}{\cellcolor[HTML]{FFFFFF}Conductance of rod cells} & \multicolumn{1}{c|}{\cellcolor[HTML]{FFFFFF}Baylor \textit{et al.} (1986)} \\ 
\multicolumn{1}{|c|}{\cellcolor[HTML]{FFFFFF}Phototransduction in rod cells} & \multicolumn{1}{c|}{\cellcolor[HTML]{FFFFFF}Forti \textit{et al.} (1989)} \\ 
\multicolumn{1}{|c|}{\cellcolor[HTML]{FFFFFF}Electrical response of cone cells} & \multicolumn{1}{c|}{\cellcolor[HTML]{FFFFFF}Baylor \textit{et al.} (1974)} \\ 
\multicolumn{1}{|c|}{\cellcolor[HTML]{FFFFFF}Ion current in bipolar cells} & \multicolumn{1}{c|}{\cellcolor[HTML]{FFFFFF}Usui \textit{et al.} (1996)} \\
\multicolumn{1}{|c|}{\cellcolor[HTML]{FFFFFF}Hyperpolarization in cell body} & \multicolumn{1}{c|}{\cellcolor[HTML]{FFFFFF}Kaneko \textit{et al.} (1985)} \\
\multicolumn{1}{|c|}{\cellcolor[HTML]{FFFFFF}Calcium current in axon} & \multicolumn{1}{c|}{\cellcolor[HTML]{FFFFFF}Tachibana  \textit{et al.} (1991)} \\
\multicolumn{1}{|c|}{\cellcolor[HTML]{FFFFFF}Calcium-dependent chloride current in cell body} & \multicolumn{1}{c|}{\cellcolor[HTML]{FFFFFF}Tachibana \textit{et al.} (1993)} \\
\multicolumn{1}{|c|}{\cellcolor[HTML]{FFFFFF}Delayed rectifying potassium current in cell body} & \multicolumn{1}{c|}{\cellcolor[HTML]{FFFFFF}Lasater (1988)} \\
\multicolumn{1}{|c|}{\cellcolor[HTML]{FFFFFF}GABA-induced current in the axon} & \multicolumn{1}{c|}{\cellcolor[HTML]{FFFFFF}Attwell \textit{et al.} (1987)} \\
\multicolumn{1}{|c|}{\cellcolor[HTML]{FFFFFF}Glutamate-induced current in the dendrite} & \multicolumn{1}{c|}{\cellcolor[HTML]{FFFFFF}Attwell \textit{et al.} (1987)} \\
\multicolumn{1}{|c|}{\cellcolor[HTML]{FFFFFF}Glutamate-induced current in the dendrite} & \multicolumn{1}{c|}{\cellcolor[HTML]{FFFFFF}Nawy \textit{et al.} (1990)} \\
\multicolumn{1}{|c|}{\cellcolor[HTML]{FFFFFF}Glutamate-induced current in the dendrite} & \multicolumn{1}{c|}{\cellcolor[HTML]{FFFFFF}Shiells \textit{et al.} (1994)} \\
\multicolumn{1}{|c|}{\cellcolor[HTML]{FFFFFF}Ion channels of AII amacrine cell} & \multicolumn{1}{c|}{\cellcolor[HTML]{FFFFFF}Smith \textit{et al.} (1995)} \\
\multicolumn{1}{|c|}{\cellcolor[HTML]{FFFFFF}Action potentials in AII amacrine cells} & \multicolumn{1}{c|}{\cellcolor[HTML]{FFFFFF}Boos \textit{et al.} (1993)} \\
\multicolumn{1}{|c|}{\cellcolor[HTML]{FFFFFF}Transient response in AII amacrine cells} & \multicolumn{1}{c|}{\cellcolor[HTML]{FFFFFF}Nelson \textit{et al.} (1982)} \\
\multicolumn{1}{|c|}{\cellcolor[HTML]{FFFFFF}Transient response in AII amacrine cells} & \multicolumn{1}{c|}
{\cellcolor[HTML]{FFFFFF}Dacheux \textit{et al.} (1986)} \\
\multicolumn{1}{|c|}{\cellcolor[HTML]{FFFFFF}Impulse encoding of ganglion cells} & \multicolumn{1}{c|}{\cellcolor[HTML]{FFFFFF}Fohlmeister \textit{et al.} (1997)} \\
\multicolumn{1}{|c|}{\cellcolor[HTML]{FFFFFF}Repetitive firing of ganglion cells} & \multicolumn{1}{c|}{\cellcolor[HTML]{FFFFFF}Fohlmeister \textit{et al.} (1990)} \\
\multicolumn{1}{|c|}{\cellcolor[HTML]{FFFFFF}Current through surface membrane} & \multicolumn{1}{c|}{\cellcolor[HTML]{FFFFFF}Hodgkin \& Huxley (1952)} \\
\multicolumn{1}{|c|}{\cellcolor[HTML]{FFFFFF}Gap junctions in dynamic range enhancement} & \multicolumn{1}{c|}{\cellcolor[HTML]{FFFFFF}Publio \textit{et al.} (2009)} \\
\multicolumn{1}{|c|}{\cellcolor[HTML]{FFFFFF}Calcium modulation in photoreceptor synapses} & \multicolumn{1}{c|}{\cellcolor[HTML]{FFFFFF}Kourennyi \textit{et al.} (2004)} \\

\hline \vspace{-6pt}
  &  \\ \hline
\end{tabular}%
\end{table}

Cascading the retina cell models from Table I results in a high-level architecture in a signal flow that looks like Fig.~\ref{fig2}. Our model also includes the option to decouple the cone and rod cell pathways, and to simulate their performance in isolation from one another. The user-configurable parameters of electrical and chemical synapses are maximum conductance, reversal potential slope (in millivolts), and the synaptic time constant. As we are concerned with one single unit of signal flow (i.e., the path of photoreceptor activation through one single neuron of each class: rod/cone, bipolar, amacrine, ganglion cells), the simulator only factors in laterally connected cells within this unit, and neglects other lateral connections. This means receptive fields and laterally-connected horizontal cells are not required in this model, though the program has been designed such that they may be included in a scaled implementation.


\section{Retina Simulator}
The simulator was developed in C++, graph plotting in Kst, and the graphical interface in Microsoft Foundation Class. It is recommended for use on Windows 10. To enable real-time processing, the numerical approaches from \cite{Eshraghian20182, Cho2016} were adopted. The numerical approach tests a range of differential equation solvers from the MATLAB suite (ODE15s, ODE45, ODE23s, ODE113), as well as the midpoint method and fourth-order Runge-Kutta (RK4) method. It was found that ODE45 and ODE113 required time steps that were far too small for real-time simulation in order to converge stable and accurate solution. ODE15s and ODE23s are both suitable for stiff systems, but the latter would only perform effectively with crude error tolerances. ODE15s proved to be the fastest and most accurate of the MATLAB solvers, but RK4 was consistently faster for fixed time steps. For the default time step of 1~$\mu$s, the RK4 method was 18.8\% faster than the ODE15s solver. As such, we adopted the RK4 method into our simulator. Optimizing the solver method was crucial, as discrete neuronal networks that cascade populations of cells are often far too computationally expensive for real-time performance \cite{Rattay2003, Resatz2004, Publio2009, Publio2012}. For quantitative results of numerical modeling of single-compartment retinal cells, we refer the reader to \cite{Eshraghian20182}.

\begin{figure}[t]
\centerline{\includegraphics[scale=0.6]{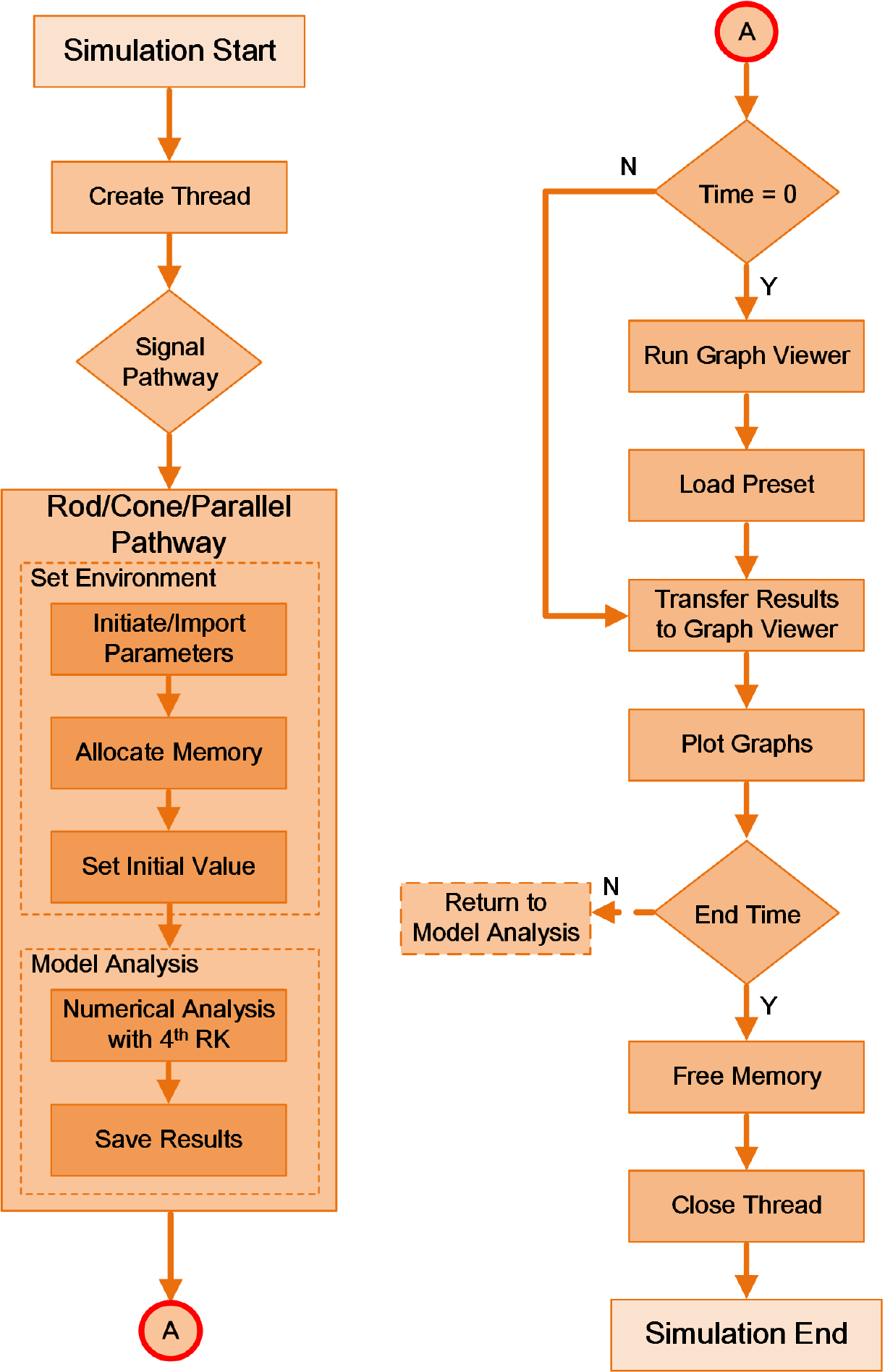}}
\caption{Simulation flow chart of the retina simulator. Left: simulation start to model analysis. Right: graph plotting to simulation end.}
\label{fig3}
\end{figure}

The simulation flow chart is shown in Fig.~\ref{fig3}, where the process for model analysis and plotting are designed to maximize CPU utilization where the number of physical threads are limited. A key specification of our simulation was accessibility to the layperson, so this enables our simulator to be run locally without the need for high-performance GPUs or CPUs. 


\begin{figure*}[t]
\centerline{\includegraphics[scale=0.4]{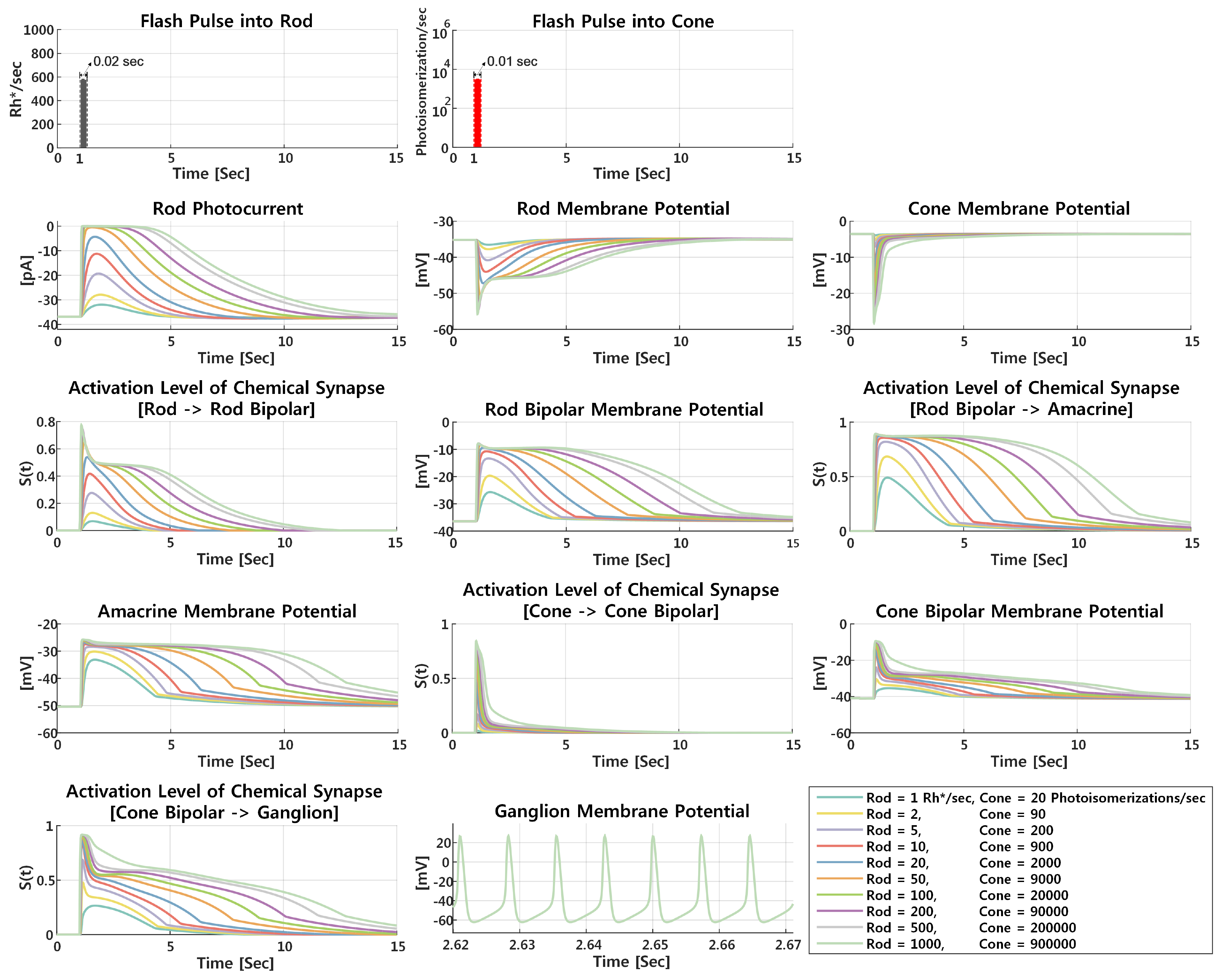}}
\caption{Simulation results of the parallel signal pathway under varying flash intensities.}
\label{fig5}
\end{figure*}

\section{Results and Discussion}
Simulation results are displayed in Fig.~\ref{fig5} across varying intensity of light injection for a run time of 15 seconds, with rod cell activations stimulated in rhodopsin per second (Rh*/sec), and cone cell activations in photoisomerizations/sec. Each stimulus is applied as a spatially uniform pulse of light. Information about the input stimulus is encoded in the number of spikes, their amplitude, shape and spike-timing, and provides a more biologically accurate representation of input to the visual processing region of the brain than block-compartment models. This is verified by comparing each individual plot from Fig.~\ref{fig5} with the patch-clamp data provided in the literature from Table~I, where the model and simulator used retains the observed biophysical properties of the retina cells. As the activation levels of the photoreceptors are increased, spikes are generated with increasing frequency, amplitude of photocurrent generated in the outer segment of the rod cell decreases, and photocurrent induces hyperpolarization in the inner segment of the rod cell. Importantly, all simulations were performed in biological real-time.

We note some of the behavioral features that can be discerned on inspection in Fig.~\ref{fig5}. Spike-frequency adaptation partially stems from the upstream neuron level, where the photoreceptor cells gradually converge back to their resting potentials upon exposure to light stimuli. The fast-spiking nature of ganglion cells is noted based on the small timescale of the ganglion membrane potential response. Subthreshold responsitivity occurs at the minimal activation of 1~Rh*/sec, although such subthreshold responses are typically filtered out in the brain to optimize for photonic noise \cite{Baylor1979}. 

The dominant model type of discrete neuronal blocks are of the single-compartment variety. An advantage of this approach is that they've been used to model nearly all neuron types, which means they are integrated together into a larger-scale neuronal network, simulated with low computational cost. However, the simplifications made in single-compartments means there are morphological features that often go ignored. Accuracy can be improved by employing multi-compartmental models \cite{Schiefer2006, Werginz2014}, but this would remove the possibility of real-time processing on limited computational resources. We partially cure this shortcoming by introducing intracellular calcium dynamics responsible for temporal spiking properties, and using optimized sodium and potassium gating kinetics which exhibits the replication of a wider range of spiking behaviors over the Hodgkin-Huxley formulation in terms of impulse encoding flexibility \cite{Fohlmeister1990, Fohlmeister1997}.

The synapse activation level is expressed as a value between 0 and 1, and contributes substantially to generating the chemical synaptic current. It approaches 1 when the membrane potential of the photoreceptor diverges from the resting state potential. Therefore, the synapse activation level increases for stronger light intensities, which facilitates higher current density transferred to subsequent cells in the neuronal cascade, causing larger or more sustained downstream cell membrane potential responses.

\section{Conclusion}
A discrete neuronal network simulator of the retina is presented in this paper. The simulator provides users with a convenient and intuitive way to simulate various dynamics of retinal cells, and the ability to reconfigure the parameters of each cell in the cascade and synaptic connection between the cells. It is expected that this simulator may provide further insight to neuroscientists and physiologists alike, in exploring the dependencies that exist between the numerous components of the retina. It may also be used by deep learning practitioners and machine learning engineers who are engaging with biologically plausible modes of learning to simplify the acquisition of retinal data, which can be used to create more accurate models of neural encoding and decoding in the visual cortex. The simulator is accessible at the following link: \url{https://github.com/sbbaek-cbnu/artificial_retina_simulator_github}. 


\end{document}